# Impact Ionization in β-Ga$_2$O$_3$

*Krishnendu Ghosh and Uttam Singisetti*
*Electrical Engineering Department, University at Buffalo, Buffalo NY 14260, USA*
*kghosh3@buffalo.edu , uttamsin@buffalo.edu*

**Abstract**

A theoretical investigation of extremely high field transport in an emerging wide-bandgap material β-Ga$_2$O$_3$ is reported from first principles. The signature high-field effect explored here is impact ionization. Interaction between a valence-band electron and an excited electron is computed from the matrix elements of a screened Coulomb operator. Maximally localized Wannier functions (MLWF) are utilized in computing the impact ionization rates. A full-band Monte Carlo (FBMC) simulation is carried out incorporating the impact ionization rates, and electron-phonon scattering rates. This work brings out valuable insights on the impact ionization coefficient (IIC) of electrons in β-Ga$_2$O$_3$. The isolation of the Γ point conduction band minimum by a significantly high energy from other satellite band pockets play a vital role in determining ionization co-efficients. IICs are calculated for electric fields ranging up to 8 MV/cm for different crystal directions. A Chynoweth fitting of the computed IICs is done to calibrate ionization models in device simulators.

## I. Introduction

Wide-bandgap semiconductors are attractive for high-power electronics, and UV optoelectronic applications. A recently emerged material β-Ga$_2$O$_3$ has gained a lot of attention [1] due to its immense potential in both electronics and photonics. High break-down voltage MOSFETs [2-6], Schottky diodes [7, 8], and deep UV photodetectors [9, 10] are experimentally demonstrated. Well-developed bulk and thin film growth techniques [11, 12] make this material a strong candidate for future applications. Accurate n-type doping and the difficulty in p-type doping make electrons the dominant carriers in this material. Electronic structure [13, 14], optical absorption [15, 16], and lattice dynamical calculations [17-19] in this material have been reported in the last few years. Theoretical investigation of electron transport in this material is crucial to augment the experimental advancements. There have been a few low-field transport calculation reports [20, 21] in this material revealing that the long-range polar optical phonon (POP) electron-phonon interactions (EPI) limit the electron mobility. Recently, the authors reported a high-field transport calculation including full-band EPI to predict velocity-field curves in this material [22] for an electric field up to 0.4 MV/cm. However, as the electric field is further increased, interband transitions and electron ionization become important. Indeed, in power devices the electric field reaches up to several MV/cm and the resulting ionization of electrons could lead to breakdown of devices. An empirical estimate of critical breakdown electric field in β-Ga$_2$O$_3$ based on bandgap was reported [1]. The authors have previously reported [23] on the impact ionization co-efficient in β-Ga$_2$O$_3$ using a simple classical calculation. But, given the exponential sensitivity of the ionization co-efficients on the electric field it is crucial to carry out a much rigorous calculation from first-principles. Proper understanding of electron-electron interaction is the key to probe impact ionization.

Recent advancement in electronic structure and lattice dynamics calculations motivates accurate model development for non-equilibrium carrier dynamics. There have been several reports on first-principles EPI calculation and subsequent carrier relaxation time estimation with high accuracy [24-26] . There are reports on the high-field transport based on empirical pseudo-potential models under rigid-ion approximations [27] that include impact ionization. Conventionally, deformation potential based theories for EPI and a Keldysh empirical ionization model are common practice [28, 29] in Monte Carlo simulations. Here, the authors explore impact ionization starting from density functional theory (DFT) under local density approximation (LDA). The uniqueness of this work is that the authors utilize maximally localized Wannier functions (MLWFs) to calculate electron-electron interactions (EEI). While MLWFs have been used in EPI formulation with high accuracy [30], they have not been used to calculate EEI to the best of the authors' knowledge. Using MLWFs in computing EEI helps integrating EPI and EEI under the same theoretical manifold thus providing a single framework for far-from-equilibrium transport calculations. Using these EEI calculations, ionization rates are obtained using Fermi-Golden rule. Full-band Monte Carlo (FBMC) simulation is carried out to extract the IICs. The IICs are fitted to a Chynoweth model to help calibrate device simulators. First, the authors discuss the theory and methods for the calculations followed by the results obtained for β-$Ga_2O_3$.

## II. Electron-phonon interactions

The primary interaction mechanism that limits transport of hot electrons is EPI. The effect of EPI is shown in a schematic on the left pane of Fig. 1. The type of EPI that controls the transport depends on the regime of the electronic transport. Under low-electric field the dominating EPI mechanism in ionic semiconductors is due to polar optical phonons (POP). The POP scattering rates are calculated using Fermi-Golden rule as follows –

$$S_{POP}(m\boldsymbol{k}) = \sum_{v,\boldsymbol{q}} w_{\boldsymbol{q}} |g_{POP}^{v}(\boldsymbol{q})|^2 \, Im\left(\frac{1}{E_{m\boldsymbol{k}+\boldsymbol{q}} - E_{m\boldsymbol{k}} \pm \omega_v^{POP}(\boldsymbol{q}) - i\delta}\right) \quad (1)$$

Here $\boldsymbol{k}, \boldsymbol{q}, m,$ and $v$ are electronic wave-vector, phonon wave-vector, and electronic band index and phonon mode index respectively. $w_{\boldsymbol{q}}$ is the weight of the $\boldsymbol{q}$ point arising from the sampling of the phonon Brillouin zone. $E_{m\boldsymbol{k}}$ is the Kohn-Sham energy eigen value of an electron in band $m$ and wave-vector $\boldsymbol{k}$, $\omega_v^{POP}(\boldsymbol{q})$ is the phonon energy eigen value of a phonon in mode $v$. $\delta$ is a small smearing in energy. $g_{POP}^{v}(\boldsymbol{q})$ is the long-range coupling elements calculated using Vogl model [31]. The Vogl model requires the Born effective charge tensor and the phonon displacement patterns both of which come from DFPT calculations. Also, note that in the current formulation $g_{POP}^{v}(\boldsymbol{q})$ is independent of the electronic wave-vector. This is because under the long-wavelength limit ($\boldsymbol{q} \to \boldsymbol{0}$) the overlap of the initial and final electronic states from the same band results to unity due to the orthogonality of the Kohn-Sham states. This is also the reason behind not considering any interband scattering mediated by POP. The electronic energies used are the Wannier interpolated energies from the actual Kohn-Sham calculation. Note that this formulation takes into account the anisotropies in the POP matrix elements arising from the low-symmetry of the crystal. For screening of the POP elements a standard Lyddane-Sachs-Teller model is used where it is assumed that a given mode could only be screened by the modes that are higher in energy compared to that mode. No screening contribution from free carriers is taken into account.

As the electric field is increased short-range non-polar phonon (NP) scatterings become important and those scatterings provide higher momentum relaxation due to the involvement of long phonon wave-vectors. The scattering rate is calculated in a similar way as in Eq. 1, however with the exception that the sum in Eq. 1 runs over final electronic band indices besides $v, \boldsymbol{q}$ since interband scatterings are possible by short-range coupling. The short-range coupling elements are computed as:

$$g_{NP}^{v,mn}(\boldsymbol{k}, \boldsymbol{q}) = \left\langle \psi_{\boldsymbol{k}+\boldsymbol{q}}^{n} \left| \frac{\partial V_{scf}^{v}}{\partial \boldsymbol{q}} \right| \psi_{\boldsymbol{k}}^{m} \right\rangle \quad (2)$$

Here, $\psi_{\boldsymbol{k}}^{m}$ s are the electronic wave-functions and $\frac{\partial V_{scf}^{v}}{\partial \boldsymbol{q}}$ represent the perturbation in the periodic crystal potential for a phonon mode $v$. The computation of $g_{NP}^{v,mn}(\boldsymbol{k}, \boldsymbol{q})$ on fine electronic and phonon BZ meshes are done using an efficient Wannier function scheme [30, 32]. Note that the overlap of the initial and final electronic states in this case cannot be approximated unlike the long-range case and needs to be explicitly taken care of. Computational details of the short-range elements in β-$Ga_2O_3$ is documented in a previous work by the current authors [22].

## III. Electron-electron interactions

Electron-electron interaction is a two particle process and the matrix element for the interaction term can be written as –

$$g^{ee}_{\substack{k_1 n \to k'_1 n' \\ k_2 m \to k'_2 m'}} = \int dr_1 \int dr_2\, \varphi^*_{k'_1 n'}(r_1)\varphi^*_{k'_2 m'}(r_2) V(r_1 - r_2)\varphi_{k_1 n}(r_1)\varphi_{k_2 m}(r_2)\, \delta(k_1 + k_2 - k'_1 - k'_2) \tag{3}$$

Here, $\varphi_{k_1 n}$ denotes the electronic wavefunction with wavevector $k_1$ and on band $n$ and $V(r_1 - r_2)$ is the screened Coulomb interaction which (in atomic units) is given by

$$V(r_1 - r_2) = \frac{e^{-q_D|r_1 - r_2|}}{\varepsilon |r_1 - r_2|} \tag{4}$$

where, $\varepsilon$ is the dynamic dielectric constant and $q_D$ is the Debye length. Writing the Coulomb potential as the sum of Fourier components and obeying the momentum conservation one can rewrite Eq. (3) as

$$g^{ee}_{\substack{n k_1 \to n' k_1 + q \\ m k_2 \to m' k_2 - q}} = \frac{4\pi}{\Omega} \frac{1}{\varepsilon(q^2 + q_D^2)} \langle n' k_1 + q | e^{iq\cdot r} | n k_1 \rangle \langle m' k_2 - q | e^{-iq\cdot r} | m k_2 \rangle \tag{5}$$

Now using maximally localized Wannier (MLW) functions [32], the Bloch electronic wave-functions can be written as $|nk\rangle = \sum_{mR_e} U_k^\dagger e^{ik\cdot R_e}|mR_e\rangle$, where $|mR_e\rangle$ is an MLW function centered at $R_e$, $U_k$ is the gauge-transforming unitary matrix that rotates the gauge of the Bloch functions in a way such that the subsequent Wannier functions achieve minimum spread in real space. Utilizing the orthonormality relation of the Wannier functions and under a small $q$ limit, one can write

$$\langle n' k_1 + q | e^{iq\cdot r} | n k_1 \rangle \big|_{q\to 0} = [U_{k_1 + q} U^\dagger_{k_1}]_{n'n} \tag{6}$$

A similar expression is used in [31] to calculate polar electron-phonon interactions in Wannier gauge. In the current context the small $q$ limit is justified due to the long-range nature of the Coulomb interactions.

The advantage of the proposed approach of using MLW functions for EEI is the ability to obtain a very fine sampling of the Brillouin zone in calculating the interaction terms. The $U_k$ elements are the eigen functions of the interpolated Kohn-Sham (KS) Hamiltonian. First the $U_k$ matrices are calculated on the coarse mesh followed by a Fourier interpolation of the KS Hamiltonian in the Wannier gauge, which will give the $U_k$ matrices on a finely sampled Brillouin zone. Impact ionization is a two electron process which involves relaxation of a hot electron in the conduction band and excitation of a valence electron into the conduction band. This is schematically shown along with a Feynman diagram on the right side pane of Fig. 1. Using the prescription given in Eq. 6, the impact ionization interaction term can be written as –

$$g^{ee}_{\substack{mk \to nk+q \\ n'k' \to m'k'-q}} = V_q\, [U_{k+q} U^\dagger_k]_{nm}\, [U_{k'-q} U^\dagger_{k'}]_{m'n'} \tag{7}$$

Here $|mk\rangle$ and $|nk + q\rangle$ are the initial states of the hot electron before and after ionization respectively and $|n'k'\rangle$ and $|m'k' - q\rangle$ are the states of the electron being ionized before and after ionization respectively. $V_q = \frac{4\pi}{\Omega \varepsilon (q^2 + q_D^2)}$ is the Fourier transformed Coulomb interaction. Screening of the Coulombic interaction is taken into account by considering a dynamic polarizability (frequency dependent) [33] under a long-wavelength limit. The ionization rate is calculated in a similar way as electron-phonon scattering rates using Fermi-Golden rule enforcing the energy conservation $\delta(E_{mk} + E_{n'k'} - E_{m'k'-q} - E_{nk+q})$. As seen on the electron-electron interaction diagram shown on Fig. 1, calculating the ionization rate for the electron at $|mk\rangle$ involves summing over the internal degrees of freedom which implies integrations over $k'$ and $q$, while summations over $m'$, $n'$, and $n$. So, someone interested in just evaluating the ionization rates need not store the matrix elements in Eq. 5, rather the rates can be computed on the fly by summing over the internal degrees of freedom. However, for final state calculation in the FBMC simulation, the entire matrix elements $g^{ee}_{\substack{nk_1 \to n'k_1 + q \\ mk_2 \to m'k_2 - q}}$ are required and hence they need to be stored. It is important to note that the $k'$ being confined to the valence band manifold only, a much coarser grid is used for the $k'$ integration since valence bands are flat. To correct the mean-field (LDA) estimated bandgap the conduction band energies are shifted to match the experimental bandgap. It is noted that the mean-field estimated wave-functions are close to the actual quasiparticle wave-functions [33], and hence the formulation of $U_k$ starting from LDA estimated wave-functions is justified. Also it is to be noted that, in the current work only the direct term of the electron-electron interaction is considered. As pointed out in [34], the contribution from the exchange term could be maximum in case point interactions and under such situation the ionization rates could be at most underestimated by a factor of 0.75 due to neglecting the exchange term. However the Coulomb interaction is far from behaving as a point interaction and hence the factor of underestimation is expected to be much lower. For electron-electron scattering rate calculations an importance-sampling (with Cauchy distribution) of the Brillouin zone is used with peak density of the **q** points at the zone center and gradually the density decays out towards the zone edges. This helps in capturing the long-range behavior of the Coulomb interaction within a reasonable computational cost. This type of sampling has been used previously for polar optical phonon scattering calculations in GaAs [24].

## IV. Ionization co-efficients from FBMC

Electrons get energized by the applied electric field and relax momentum and energy through EPI and EEI. Under low-field this could be treated by relaxation time approximation or iterative techniques starting from equilibrium distribution. However, high-field drives the distribution far away from equilibrium and the low-field techniques fail. Monte Carlo simulation stochastically compute the trajectories of the ensemble of electrons. The details of the simulation can be found [22]. EEI is added as an additional scattering mechanism in a way similar to EPI. The EEI is restricted to be only between valence electrons and conduction electrons. In other words EEI within the conduction band manifold is not considered and that does not affect the computed transport properties since the ensemble averages remain unchanged by mere exchange of momentum and energy within the conduction band. The valence electrons getting excited to the conduction band are referred as secondary electrons and the high energy conduction band electron losing energy as the primary electron. As described in the previous work [22] by the authors, in case of EPI mediated scatterings, during the final state selection in the FBMC algorithm the scattering mechanisms were classified based on final band index of the scattered electron, phonon mode index, polar/non-polar nature of the scattering and absorption/emission. In case of EEI, the corresponding mechanisms are classified as the conduction band index of the primary and secondary electrons after ionization and the **k**-point index of the secondary electron before ionization (which is defined on a much coarser grid than the actual fine **k** grid, see sec. III). The generation rate of the secondary electrons (G) is computed to be as the difference of the total number of electrons at the end of the simulation to that at the beginning divided by the simulation time. The IIC ($\alpha$) is defined as the reciprocal of the mean free path traversed by an electron before creating an ionization. The IIC is extracted from the generation rate using the relation $G(F) = \alpha(F)v_d(F)$ where $v_d(F)$ is the drift velocity for an applied electric field F which is calculated in the FBMC simulation.

## V. Results and discussions

First, density functional theory (DFT) calculations are carried on $\beta$-Ga$_2$O$_3$ unit cell [13] under LDA using norm-conserving pseudopotentials [35] in Quantum ESPRESSO [36]. The Brillouin zone is sampled with Monkhorst-Pack [37] grid of 8×8 ×4 with an energy cut-off 80 Ry to truncate the reciprocal vectors. The Kohn-Sham eigen values are interpolated on a fine electronic grid and the rotation matrices are computed on the same grid. The $\beta$-Ga$_2$O$_3$ conventional unit cell is shown in Fig. 2(a) (visualized by Vesta[38] ) and the interpolated KS eigen values are shown in Fig 1(b) for two reciprocal crystal directions. The inset in Fig. 2(b) shows the BZ (visualized by XCrySDen [39]) with the corresponding reciprocal directions. To obtain the screening element, $\varepsilon(\omega)$, the full-frequency epsilon calculation is used as implemented in BerkeleyGW [33] . While the details of EPI could be found in previous reports [20, 22] by the current authors, here the methodology is described in a few sentences. Using density functional perturbation theory [40], the phonon eigen values, displacement patterns, and EPI elements are calculated on coarse mesh. Next, the Wannier-Fourier interpolation [30, 32] is carried out to calculate the EPI elements and the phonon dynamical matrices. Long-range POP scattering is calculated separately following [31].

The ionization rates are computed using the theory described in Sec. III. There are a total of 6 conductions bands taken in the transport calculation and the electron-electron self-energies for conduction bands 5 and 6 are shown in Fig. 2(c) along the two reciprocal vector directions. In the Monte Carlo scheme ionization rates are included only from bands 5 and 6 since the lower bands (1-4) do not have energy states high enough to cause ionization. It is noted that band 4 has some states away from the zone center that have energy higher than the bandgap but their computed ionization rates (not shown) are much smaller than the EPI scattering rates. It could be seen that the ionization rates are much higher near the zone center compared to the zone edges. This peculiarity arises from the isolation of the conduction band minimum at the Γ point by a significantly high energy difference. To be more specific, the electrons near the zone edge in bands 5 and 6 which have energies higher than the bandgap cannot ionize as there are no available final states. The long-range nature of the interaction prohibits a zone edge electron to release enough energy (> bandgap) and end up near the zone center. Hence the ionization rates mediated by the zone edge hot electrons are significantly lower than that by zone center electrons. No phonon-assisted ionization mechanism is considered here, however. Fig. 3(a) shows the contributions of the individual electronic bands in impact ionization. It could be seen that qualitatively the ionization rate follows a power law [29] near the ionization threshold which is slightly above 5 eV. The abrupt drop of the ionization rates around 6.5 eV does not happen due to truncation of the conduction bands after the sixth band. There are significant available states on the sixth band much beyond 6.5 eV as well. The abrupt drop rather arises from the absence of the final density of states after an ionization event. Since the bandgap is around 4.9 eV, the final energy of the ionizing electron has to be at least 4.9 eV less than the energy it had prior to ionization. However, due to the absence of remote satellite valleys within 0-2.4 eV of the CB minima the hot electrons can only end up on the Γ valley. On the sixth band the states which possess more than 6.5 eV occur much away from the Γ point and hence lacks final density of states to cause an ionizing event.

The best practice to determine the number of conduction bands to be taken in the Monte Carlo simulation should be 'a posteriori'. However, there are a few important aspects to be noted here. The hot electron distribution function dies off quickly after its peak and only a minuscule of the population survives beyond 6eV as is shown on Fig. 3(b). This is the scenario even for the highest magnitude of the electric field considered in this work which is 8 MV/cm. So the authors speculate that considering beyond 6 bands might not make any difference in the distribution function and hence the ionization co-efficient will not get affected. The top 4 valence bands were used in the ionization rate calculation considering a 1 eV window from the VB maxima. This translates to around a minimum of 6 eV energy for the ionizing electron. To further address the need for an 'a posteriori' convergence study with respect to the number of conduction bands, the authors would like to point out the computational limitation caused by the huge memory requirement of storing the EPI elements. The 30 phonon modes along with fine Brillouin zone sampling require an enormous amount of memory. Some programing strategies [22] are taken to circumvent the memory overflow challenges and that is why it was possible to carry out the computation with 6 bands. However, considering beyond 6 conduction bands could not be achieved. Please note that the required memory for storing the matrix elements will scale quadratically with the number of bands (since the number of allowed transitions will scale quadratically).

The FBMC scheme initializes the ensemble of electrons thermodynamically after which the electric field is turned on. Trajectories of the electrons are formed in reciprocal space stochastically by using the electron-phonon scattering rates, and electron ionization rates. Six conduction bands are taken into account in the FBMC simulation. The FBMC simulation is run or electric fields ranging from 1 MV/cm to 8 MV/cm. Fig. 4(a) shows the transient electron dynamics under an applied electric field of 2 MV/cm. The oscillations that are observed initially result when the electrons cross the Bragg planes and they are subsequently suppressed out due to EPI. Under a high enough electric field the transit time to reach the Bragg plane (let's call it $\tau_B$) can become comparable to mean free time ($\tau_s$) between successive scattering events ($\tau_B \approx \tau_s$) making it possible for the electrons to reach the plane. This oscillation, often known as Bloch oscillation (BO), is experimentally observed in superlattices [41] at room temperature, but very rare in bulk semiconductors. In β-Ga$_2$O$_3$, the satellite valleys occur at an energy comparable with the zone edge maxima of the first conduction band. Hence the onset of intervalley scattering occurs only near the zone edge where the reflection (reversal of electronic group velocity) is also likely to onset. The time-period of oscillation (TB) in Fig. 4(a) is comparable with the analytically calculated BO time-period, $T_B = \frac{\hbar}{eFd}$, where d is the distance of the Bragg plane from the zone-center further confirming the origin of the oscillations. In bulk semiconductors, there are primarily two prohibitive actions that prevent the observation of Bloch oscillations. First is the very high scattering rates mediated by non-polar optical phonons at high electronic wave-vectors demanding for an unrealistically high electric field to satisfy the criterion τ$_B$≈ τ$_s$. Secondly, under such a high electric field in most semiconductors band-to-band tunneling comes into action. This significantly increases the net number of electrons having a positive group velocity compared to the ones having negative a negative velocity (arising due to reflection on the Bragg plane) thereby prohibiting the observation of Bloch oscillation. In the current work, it is shown that it is possible the satisfy the first criterion in β-Ga$_2$O$_3$. But no band to band tunneling phenomenon is considered and hence it cannot be conclusively said whether Bloch oscillation can actually be observed.

Fig. 4(b) shows the occupation of the bands as the electric field is increased in two different directions. In this calculation the interband transitions occur only via short-range EPI and long-range EEI. It could be seen that the population on the first band drops to about half of the total as the field reaches 8 MV/cm. Although the population on bands 5 and 6 are really low even at very high fields, given the fact that impact ionization is a cumulative process, even a small population can eventually trigger avalanche breakdown. Fig. 4(c) shows the calculated IICs along three different directions. The anisotropy in the IIC is attributed to the anisotropy in EEI which in turn originates from the anisotropy of the higher conduction bands ( bands 5 and 6 in this case) even near the Γ point as seen in Fig. 2(c). There are some uncertainties with the bandgap of β-Ga$_2$O$_3$ and a range of values 4.5eV-4.9eV could be found in literature based on experiments. Hence the calculations are carried out for two different gaps 4.5eV and 4.9 eV. The uncertainty in IIC due to the bandgap uncertainty is within an order of magnitude. Given the classical exponential dependence [42] of IIC on bandgap, this much uncertainty in the IIC is reasonable. Also, as expected, the IIC is lower for a higher bandgap. For comparison, the calculated IIC is less than that computed in GaN [27] for a similar range of electric fields which indicates a higher avalanche breakdown field for β-Ga$_2$O$_3$.

Finally, to help facilitate device simulation, the authors perform a Chynoweth fitting of the calculated IIC and attempt to identify favorable transport directions for high power applications. The Chynoweth model [43] formulates the IIC as, $\alpha(F) = ae^{-\frac{b}{F}}$, where $a$ and $b$ are the fitting parameters. In order to account for the anisotropy in IIC, three sets of the parameters are provided for three different directions on Table-I. In electronic devices under high electric fields, impact ionization induced avalanche breakdown occurs if the generation of secondary carriers become self-sustainable which in turn requires the ionization integral [44] to be greater than 1. To have a qualitative understanding on the avalanche breakdown field in β-Ga$_2$O$_3$ a hypothetical triangular electric field profile (which is approximately the case in p+-n junctions) of peak electric field 8 MV/cm is considered and the base of the triangle to be 1 μm wide. Considering the estimated Chynoweth parameters, the ionization integral values are also tabulated on Table-I. The computation of ionization integral (I$_α$) requires knowledge of the ionization co-efficient of both electrons (α$_n$) and holes (α$_p$). However, the hole ionization coefficients for β-Ga$_2$O$_3$ have not been estimated. So, the ionization integral is shown for the two limits α$_n$≈α$_p$, and α$_p$<< α$_n$. As could be seen, for the considered electric field profile, avalanche breakdown is very likely on *y* direction while very unlikely on *x* direction. So for high voltage/power applications electron transport along the *x* direction is more favorable than the other two directions. Based on the hypothetical electric field profile, the critical electric fields are evaluated under the limit α$_n$≈α$_p$ and is also listed on Table-I. As revealed in a previous work [45] by the current authors and also from experimental observations [46] the electron mobility is slightly lower in the *x* direction compared to that in *y*. But considering the joint effect of on-resistance and breakdown field, transport along *x* direction is supposed to provide a higher Baliga's figure of merit [47]. Hence one can conclude that *x* direction is the most suitable transport direction for high power applications using β-$Ga_2O_3$.

## VI. Conclusions

Maximally localized Wannier functions are utilized to compute the ionization rates in β-$Ga_2O_3$ from first-principles. FBMC simulation is done to compute IICs for a wide-range of high electric fields along two different directions. IIC is fitted to an empirical Chynoweth model to calibrate device simulators. A hypothetical estimate using the computed Chynoweth parameters predicts avalanche breakdown field to be higher than the empirically predicted value of 8 MV/cm in the *x* direction.

The authors acknowledge the support from the National Science Foundation (NSF) grant (ECCS 1607833) monitored by Dr. Dimitris Pavilidis. The authors also acknowledge the excellent high performance computing cluster provided by the Center for Computational Research (CCR) at the University at Buffalo.

**Figures and Tables**

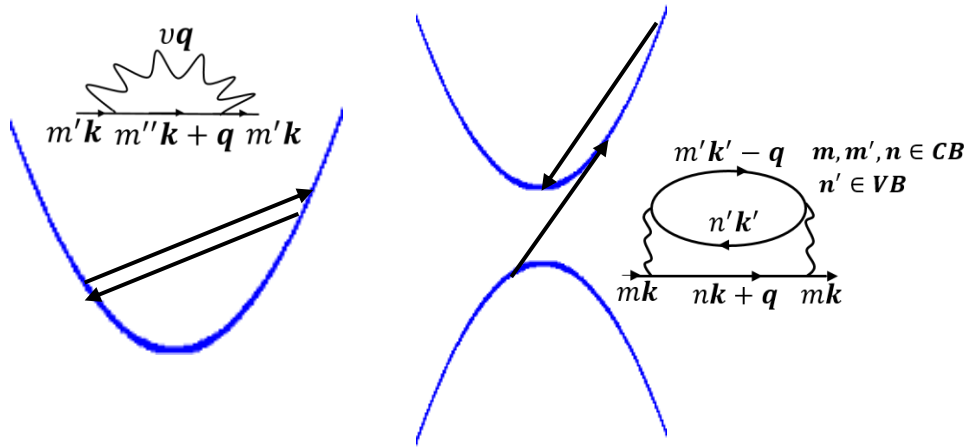

Fig. 1: Schematic showing the EPI event considered in this work, (right pane) Electron-electron interaction is shown as a two-electron process leading to the formation of an electron-hole pair. Corresponding Feynman diagram is also shown.

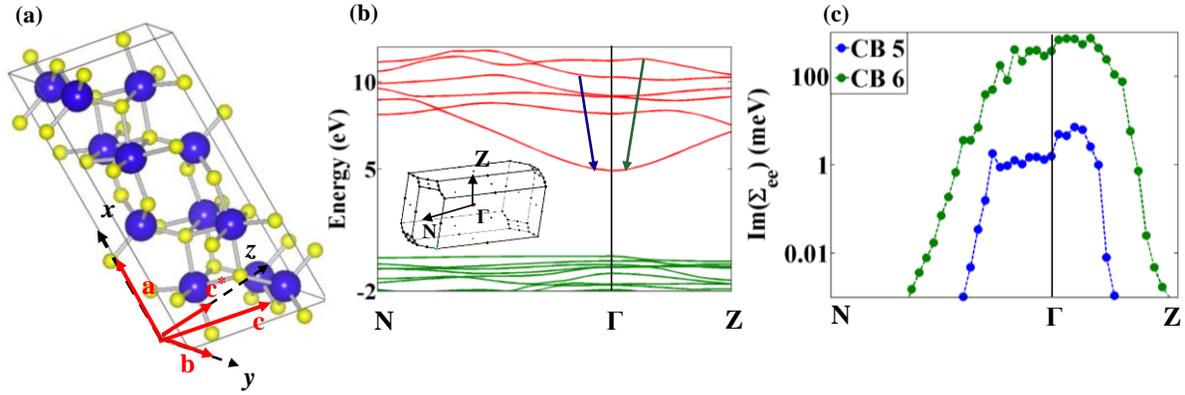

Fig. 2: The monoclinic lattice with the Cartesian directions used in this work. The crystal lattice directions are also shown. The angle between c and c* is $13.83^0$. Larger atoms are Ga while smaller ones are O (b) The Wannier interpolated band-structure in two reciprocal directions. The mean-field computed excited-state eigen values are scissor shifted to match experimental bandgap. The blue and green arrows show the two possible EEI mediated transitions taken in the calculation. (inset) the first BZ. (c) The imaginary part of the computed electron-electron self-energy using MLWF and Fermi-Golden rule for two conduction bands.

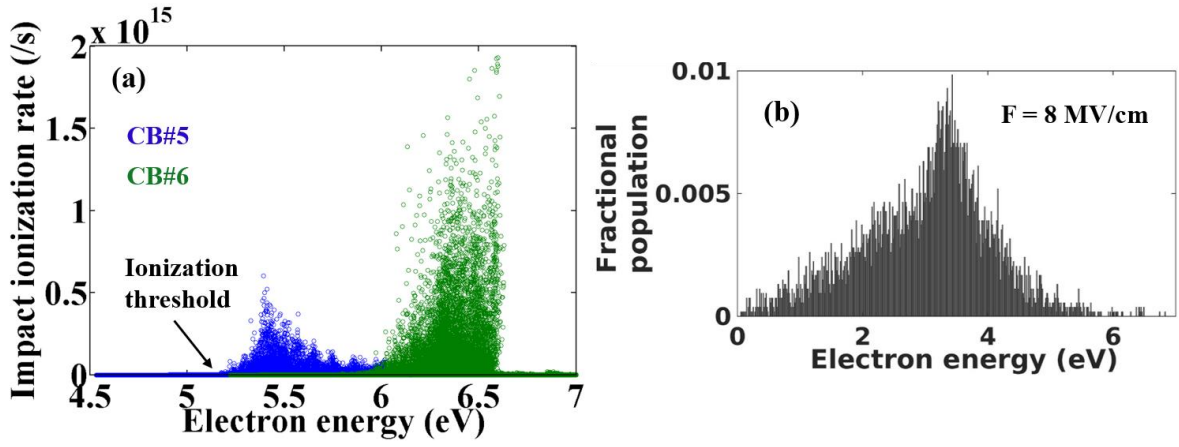

Fig. 3: (a) Contribution of the individual bands to impact ionization. (b) Distribution function of the hot electrons at an electric field of 8MV/cm

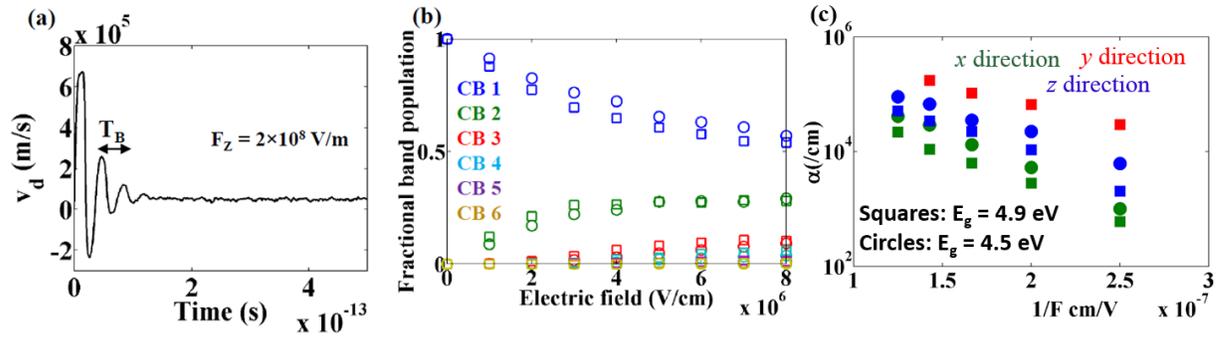

Fig. 4: (a) The Bloch oscillations under an electric field $2\times10^8$ V/m. The time period of the oscillation matches well with the analytical BO time period (see text for details). (b) The fractional band population as a function of the electric field from FBMC simulation. Circles represent the case when the applied electric field is in *x* direction, while squares represent the same for *z* direction. (c) Computed IIC for three different Cartesian directions. Due to uncertainty in the bandgap of β-$Ga_2O_3$, the IIC values are also computed for a bandgap of 4.5 eV along *x* and *z* directions (see text for details).

Table-1: The Chynoweth parameters, ionization integrals, and critical electric field values for three different Cartesian directions for a given hypothetical electric field profile (see text for details). These parameters are computed considering the bandgap to be 4.9 eV. A lower bandgap will result to a lower critical field.

| | $a$ (/cm) | $b$ (V/cm) | $I_\alpha\|_{\alpha_n\approx\alpha_p,\ E_p=8MV/cm}$ | $I_\alpha\|_{\alpha_n\gg\alpha_p,\ E_p=8MV/cm}$ | $E_c\|_{\alpha_n\approx\alpha_p}$ (MV/cm) |
|---|---|---|---|---|---|
| $x$ | $0.79\times10^6$ | $2.92\times10^7$ | 0.38 | 0.32 | 10.2 |
| $y$ | $2.16\times10^6$ | $1.77\times10^7$ | breaks down | breaks down | 4.8 |
| $z$ | $0.706\times10^6$ | $2.10\times10^7$ | breaks down | 0.70 | 7.6 |